\documentclass[secnumarabic,aps,prb,reprint]{revtex4-2}
\usepackage[english]{babel}
\usepackage{graphicx}
\usepackage{color}
\usepackage{amsmath}
\usepackage{tabularx}
\usepackage[dvipsnames]{xcolor}

\begin{document}
\title{Symmetries of electron interactions in Hubbard models of unconventional superconductors}
	\author{Sergei Urazhdin}
	\email{sergei.urazhdin@emory.edu}
	\affiliation{Department of Physics, Emory University, Atlanta, USA}
	\author{Yiou Zhang}
	\affiliation{Department of Physics, Emory University, Atlanta, USA}
	\date{\today}
\begin{abstract}

We use symmetry arguments to show that the matrix elements of electron-electron  interaction on a lattice reach extrema in states composed of wavevectors near high-symmetry points of the Brillouin zone. The mechanism is illustrated by minimal  models of cuprates and Fe-based superconductors, where this dependence originates from the wavevector-dependent orbital composition of wavefunctions. We discuss how these dependences can facilitate  finite-momentum pairing. Our results provide symmetry-based guidance for the search for new high-temperature superconductors.

\end{abstract}
	
\maketitle

\section{Introduction}\label{sec:intro}

Crystal potential of materials defines their band structure, whose importance for the correlation phenomena driven by electron interactions is widely recognized. Narrow bands tend to be unstable with respect to interactions, leading to collective electron states exemplified by superconductivity and magnetism~\cite{Balents2020-gp}. However, the effects of crystal potential on interactions are less explored.

In the conventional Bardeen-Cooper-Schrieffer (BCS) theory of superconductivity (SC), Cooper pairing results from the dynamic overscreening of Coulomb repulsion by the ions, which is well described by the continuous-medium approximation where the discrete nature of the lattice is not essential~\cite{tinkham2004introduction}. In contrast, in unconventional superconductors electron hopping is generally small, and wavefunctions are dominated by atomic components not captured by the nearly free electron approximation. 

The complexity of this problem is reflected by the diversity of the proposed mechanisms of unconventional SC, including antiferromagnetic (AF) fluctuations~\cite{doi:10.1126/science.235.4793.1196,Moriya1990}, the Kohn-Luttinger overscreening mechanism~\cite{10.1063/1.4818400,PIMENOV2022169049}, coupling of multiple electrons~\cite{Wu2024}, multi-band effects~\cite{PhysRevLett.121.187003}, hopping-induced effective attraction~\cite{doi:10.1126/science.235.4793.1196,Crpel2021}, and excitonic effects~\cite{PhysRevB.105.094506}. Since these theories have not yet achieved predictive power, analysis of the general symmetries of interactions in unconventional superconductors can provide simple guidance for the search for new superconducting materials. 

Here, we use arguments based on the lattice symmetry to analyze  the dependence of the interaction matrix elements on electron wavevectors, and identify their combinations providing interaction extrema. We illustrate the mechanisms leading to these symmetries using minimal models of cuprate and Fe-based superconductors, and discuss the forms of pairing they can facilitate.

\section{Symmetries of electron interactions on the lattice}\label{sec:lattice}

\begin{figure}
	\centering
	\includegraphics[width=0.9\columnwidth]{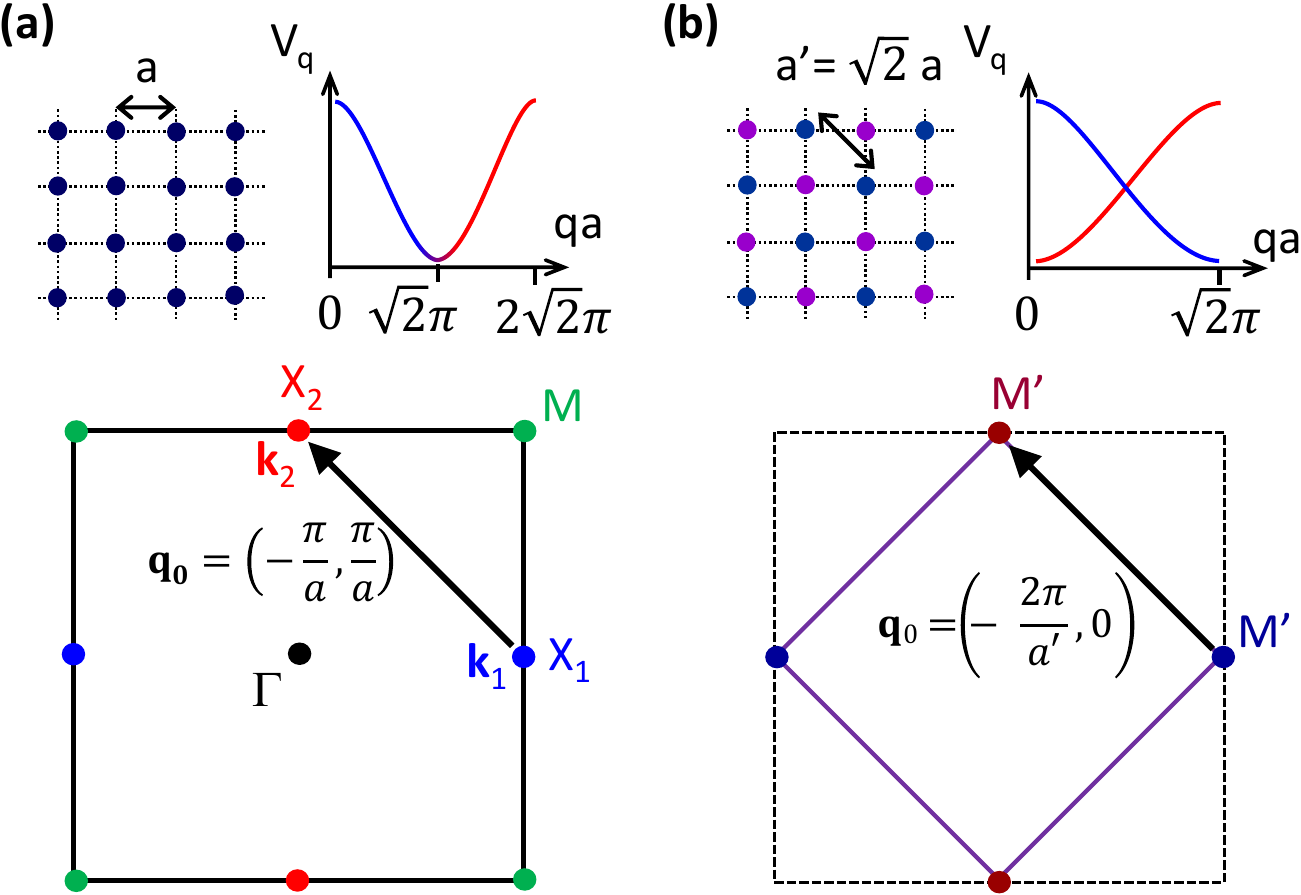}
	\vspace{-5pt}
	\caption{\label{fig:Umklapp} (a) Main panel: Momentum transfer $\mathbf{q}$ between two inequivalent $X$-points of BZ for a simple square lattice (top left inset). Top right inset: dependence of interaction on the momentum transfer along the $X_1$-$X_2$ direction. (b) Same as (a), for the  folded BZ of the bipartite square lattice.}
\end{figure}

We consider the symmetries of electron-electron interaction on the lattice, for opposite spins relevant to singlet pairing. The interaction matrix element
$$
V_{\mathbf{q},\mathbf{k},\mathbf{k}'}=\langle0|\hat{c}_{\mathbf{k}'-\mathbf{q},\downarrow}\hat{c}_{\mathbf{k}+\mathbf{q},\uparrow}\hat{H}_{int}\hat{c}^+_{\mathbf{k},\uparrow}\hat{c}^+_{\mathbf{k}',\downarrow}|0\rangle
$$
describes scattering from the states with wave vectors $\mathbf{k}$, $\mathbf{k}'$ and spins $s=\uparrow,\downarrow$ into the states $\mathbf{k}+\mathbf{q}$, $\mathbf{k}'-\mathbf{q}$. Here, $\hat{c}_{\mathbf{k},s}$ is the particle operator with wave vector $\mathbf{k}$ and spin $s$, and $|0\rangle$ is vacuum. We assume that the considered wavevectors are close to the Fermi surface, which is tailored by doping or bandstructure tuning via composition.

For the screened Coulomb potential in a translation-invariant system, the matrix element depends only on the transferred wave vector,
\begin{equation}\label{eq:vs_q}
V_\mathbf{q}=\frac{4\pi e^2}{q^2+k_s^2}\,(3D),\, V_\mathbf{q}=\frac{2\pi e^2}{\sqrt{q^2+k_s^2}}\,(2D),
\end{equation}
where $k_s$ is the Fermi screening wave number. This dependence monotonically decreases with increasing $|\mathbf{q}|$ while remaining positive. Superconducting pairing requires negative $V_{\mathbf{q}}$, which in the BCS theory results from the retarded screening by the lattice. 

Electron quasi-momentum is conserved only up to the reciprocal lattice constant. For electron-electron scattering, the matrix element $V_{\mathbf{q},\mathbf{k},\mathbf{k}'}$ is periodic with respect to electron quasi-momenta, and consequently the quasi-momentum transfer $\mathbf{q}$. Thus, if $V_\mathbf{q}$ is maximized at $\mathbf{q}=0$, it is also maximized at $\mathbf{q}=\mathbf{K}$, where $\mathbf{K}$ is a reciprocal lattice vector.

Since $\mathbf{K}/2+\mathbf{q}\equiv\mathbf{q}-\mathbf{K}/2$, in the presence of spatial inversion or time reversal symmetry the points $\mathbf{q}=\mathbf{K}/2$ are extrema of $V(\mathbf{q})$. For nearly free electrons experiencing Coulomb repulsion, the dependence $V(\mathbf{q})$ is predominantly determined by the magnitude of $\mathbf{q}$. The matrix element is then minimized by the largest irreducible wave vector,  $\mathbf{q}_0=(\pm\pi/a,\pm\pi/a)$ for the square lattice with lattice constant $a$.

Lattice potential also results in the dependence on the electron wave vectors $\mathbf{k}$, $\mathbf{k}'$, as will be illustrated by the examples in the next two sections. At extrema, $\partial{V}/\partial\mathbf{k}=\partial{V}/\partial\mathbf{k}'=0$. For $\mathbf{q}=\mathbf{q}_0$, this condition is satisfied by the pairs of high-symmetry points: the $\Gamma$-point and the $M$-point, the two inequivalent $X$-points at the BZ boundary, or for $\mathbf{k}=\mathbf{k}'$ at these points. 

We prove this for a representative example $\mathbf{k}=(\pi/a,0)\equiv\mathbf{k}_1$, $\mathbf{k}'=(0,\pi/a)\equiv\mathbf{k}_2$, the two $X$-points as shown in Fig.~\ref{fig:Umklapp}(a). Consider $V_\mathbf{\kappa}=V_{\mathbf{q}_0,\mathbf{k}_1+\mathbf{\kappa},\mathbf{k}_2}$ with a small displacement $\mathbf{\kappa}$. By the mirror symmetry with respect to the y-axis, $V_\mathbf{\kappa}=V_{\mathbf{q}_0,\mathbf{k}_2-\mathbf{q}_0,-\mathbf{k}_1-\mathbf{\kappa},\mathbf{k}_2}$, where we used $(\pi/a,\pi/a)\equiv(-\pi/a,\pi/a)$. Translating by $\mathbf{K}=(\pi/a,0)$, we obtain
$V_\mathbf{\kappa}=V_{\mathbf{q}_0,\mathbf{k}_1-\mathbf{\kappa},\mathbf{k}_2}=V_{-\mathbf{\kappa}}$. Thus, the point $\mathbf{k}=\mathbf{k}_1$ is a stationary point of $V$. 

For Coulomb repulsion in a single band, $V_{\mathbf{q},\mathbf{k},\mathbf{k}'}$ has a maximum at the $\Gamma$-point $\mathbf{q}=\mathbf{k}=\mathbf{k}=0$ corresponding to the maximum Bloch state overlap. Using the extreme value theorem for the principal directions of BZ, $\mathbf{k}=\mathbf{k}_1$, $\mathbf{k}'=\mathbf{k}_2$ are minima or saddle points. Thus, the interaction energy can be minimized at the two inequivalent $X$-points of BZ. This is realized in the model of cuprates discussed in the next section.

We now consider a bipartite square lattice where the neighboring sites are slightly different so that its unit cell is doubled, as illustrated in the top left inset of Fig.~\ref{fig:Umklapp}(b). This is realized in Fe-based superconductors due to the slightly different crystal fields of nearest-neighbor Fe atoms. The two original inequivalent X-points become equivalent $M'$-points of the folded BZ, Fig.~\ref{fig:Umklapp}(b).  
For the pairs of states that belong to the same band, the Coulomb repulsion energy is maximized at $\mathbf{q}=0$ and minimized at $\mathbf{q}_0=(\frac{\sqrt{2}\pi}{a},0)$, blue curve in the top right inset of Fig.~\ref{fig:Umklapp}(b). The opposite is true for the matrix elements between different bands [red curve]. Consequently, $\mathbf{q}_0$ is both a minimum and a maximum of interaction matrix elements, depending on the combinations of band states. The minimum can be interpreted as a consequence of umklapp scattering that maximally separates nearby electrons, minimizing their interaction energy. We confirm these properties for the model of Fe-based superconductors in Section~\ref{sec:bipartite}.

Based on these symmetry arguments, the high-symmetry points of BZ, the $\Gamma$-point $-$ the two inequivalent $X$-points and/or  the $M$-point for the square lattice $-$ are the stationary points of electron-electron interaction matrix elements with respect to the electron wavevectors and momentum transfer. Thus, to identify the possible mechanisms of attractive interaction and unconvenitonal pairing, it can be sufficient to focus on the combinations of wavevectors near these points. A corollary to this conclusion is that the materials with the Fermi surface pockets concentrated close to these points are more likely to exhibit attractive interactions. While the specific combinations of these high-symmetry points depend on the pairing mechanism, this conclusion is general and can provide useful guidance in the search for new unconventional superconductors.

\begin{figure}
	\centering
	\includegraphics[width=0.95\columnwidth]{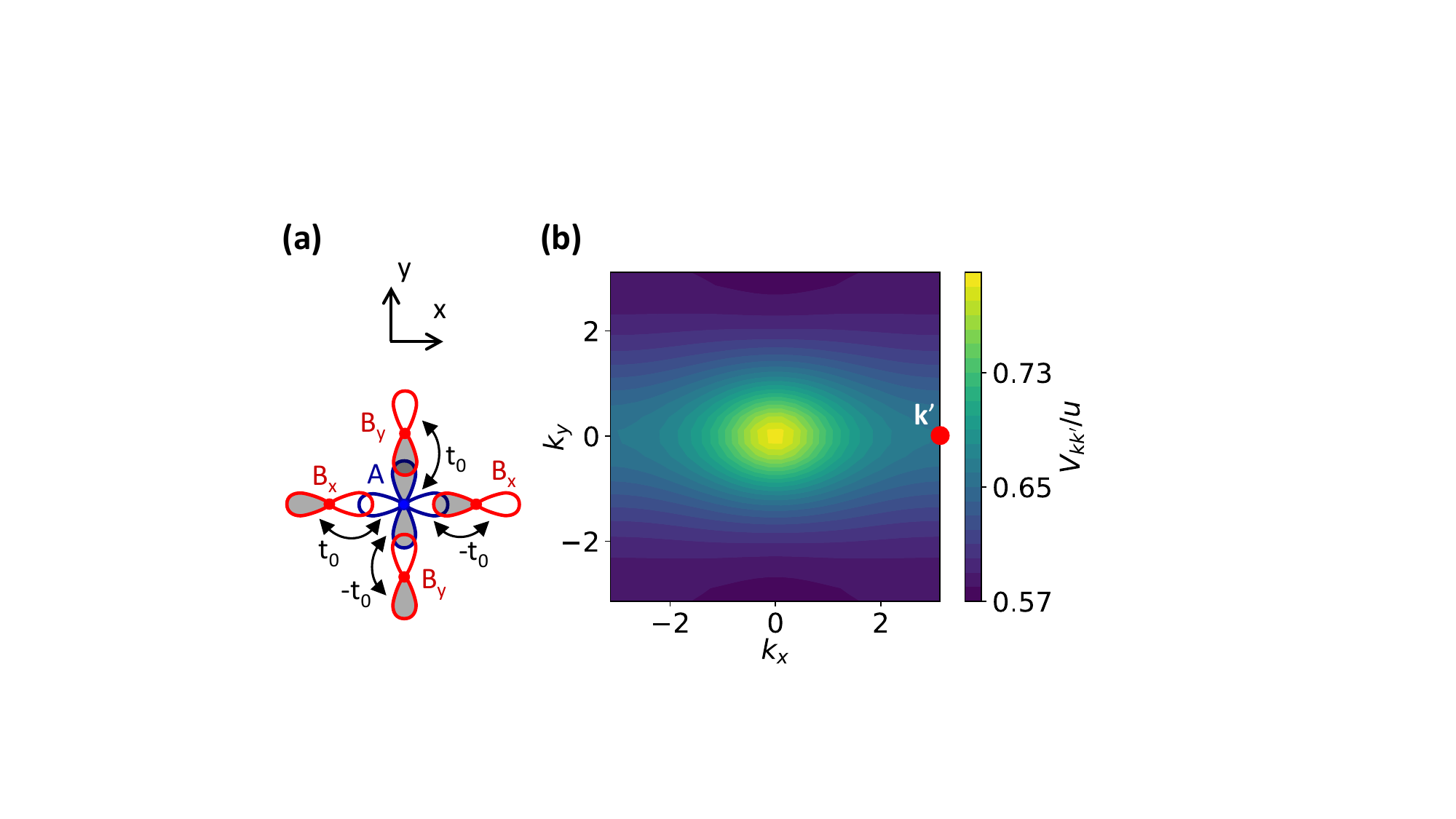}
	\vspace{-5pt}
	\caption{\label{fig:CuO} (a) The A-site and the four nearest-neighbor $B$-sites for the lattice with the AB$_2$ basis modeling the CuO$_2$ plane of cuprates. (b)  $V_{\mathbf{k},\mathbf{k}',\mathbf{q}}/u$ vs $\mathbf{k}$, for $\mathbf{k}'=\mathbf{k}_1= (\pi,0)$, $\mathbf{q}=0$, $u'=0.1u$, $t_0=0.5\epsilon_d$.}
\end{figure}

\section{Interactions on the square lattice with basis}\label{sec:cuprates}

We illustrate the symmetries discussed in the previous section for the square lattice with the basis AB$_2$ modelling the CuO$_2$ plane of cuprates.
The minimal Hubbard Hamiltonian enabling analysis of the wave vector dependence of interaction is~\cite{PhysRevB.37.3759}
\begin{equation}\label{eq:Hk_CuO}\nonumber
	\hat{H}_k=-\sum_{\mathbf{n},\mathbf{l},s} (t_\mathbf{l}\hat{d}_{\mathbf{n},s}^+\hat{p}_{\mathbf{n}+\mathbf{l}/2,s}+h.c.)
	+\sum_{\mathbf{n},s}\epsilon_d\hat{d}^+_{\mathbf{n},s}\hat{d}_{\mathbf{n},s},
\end{equation}
where $\mathbf{l}$ is a unit vector in one of the four principal directions, $t_\mathbf{l}=i^{l_x-l_y+1}t_0$,
$\mathbf{n}$ ($\mathbf{n}+\mathbf{l}/2$) enumerates lattice sites $A$ ($B$), $\hat{d}$, $\hat{p}$ are particle operators on the corresponding sites, and $\epsilon_d$ is the energy of level $A$ relative to $B$. 
This model does not include the next-neighbor hopping, which is  necessary to reproduce the electron necks at the $X$-points but does not significantly affect the orbital composition of single-particle states. 

The conduction band is derived from the site $A$ orbital, with the other two bands filled. The amplitudes $\alpha_{\boldsymbol{k}}$, $\beta_{x(y),\boldsymbol{k}}$ on sites $A$ and two inequivalent sites $B_x$, $B_y$ [see Fig.~\ref{fig:CuO}(a)] are
\begin{equation}\label{eq:ampl}\nonumber
		\alpha_\mathbf{k}=\left[2-\frac{\epsilon_d}{E_\mathbf{k}}\right]^{-1/2},
	\,\beta_{x(y),\mathbf{k}}=\pm \frac{t_0\alpha_\mathbf{k}}{E_\mathbf{k}}[e^{ik_{x(y)}}-1],
\end{equation}
where $E_\mathbf{k}=\frac{\epsilon_d}{2}+\sqrt{\frac{\epsilon_d^2}{4}+4t_0^2(\sin^2\frac{k_x}{2}+\sin^2\frac{k_y}{2})}$ is the band energy, and $a=1$ in the chosen units of length. The amplitude on sites $B_x$ ($B_y$) is maximized at the point $X_1$ ($X_2$), and vanishes at $X_2$ ($X_1$). Consequently, the interaction energy is minimized near different $X$-points due to two contributions: i) the wavefunctions do not overlap on sites $B$, and ii) because of large amplitudes on sites B, the amplitude on sites A is reduced. Using $t_\sigma/\epsilon_d\approx 0.2-0.5$ obtained from the tight-binding fitting of dispersion in cuprates~\cite{Weber2012,Kowalski2021}, we estimate that at the $X$-points the amplitudes on oxygen are comparable to those on Cu, resulting in a large difference between the orbital compositions at the two X-points.

We use the extended Hubbard interaction Hamiltonian 
\begin{equation}\label{eq:Hint_CuO}\nonumber
	\begin{split}
		\hat{H}_{int}&=U\sum_{\mathbf{n}}\hat{n}_{a,\mathbf{n},\uparrow}\hat{n}_{a,\mathbf{n},\downarrow}
		+\frac{U}{2}\sum_{\mathbf{n},\mathbf{l}}\hat{n}_{b,\mathbf{n}+\mathbf{l}/2,\uparrow}\hat{n}_{b,\mathbf{n}+\mathbf{l}/2,\downarrow}\\
		&+\frac{U'}{2}\sum_{\mathbf{n},\mathbf{l},s}\hat{n}_{a,\mathbf{n},s}\hat{n}_{b,\mathbf{n}+\mathbf{l}/2,-s},
	\end{split}
\end{equation}
where $\hat{n}$ are the corresponding particle density operators, the same Mott parameter $U$ is used for simplicity for both sites A and B, and the last term accounts for the nonlocal interaction between electrons on sites $A$ and the nearest-neighbor sites $B$. If the Mott parameter on site $B$ is significantly smaller than on site $A$, the wavevector dependence is reduced but remains non-negligible due to the variation of amplitudes on site $A$.

The matrix element is the sum of the local (onsite) and the nonlocal (nearest-neighbor) contributions, $V_{\mathbf{q},\mathbf{k},\mathbf{k}'}=V^{l}_{\mathbf{q},\mathbf{k},\mathbf{k}'}+V^{n}_{\mathbf{q},\mathbf{k},\mathbf{k}'}$. The local contribution is
\begin{equation}\label{eq:Vl_CuO2}
\begin{split}
V^{l}_{\mathbf{q},\mathbf{k},\mathbf{k}'}&=u(\alpha_{\mathbf{k}+\mathbf{q}}\alpha_{\mathbf{k}'-\mathbf{q}}\alpha_{\mathbf{k}}\alpha_{\mathbf{k}'}+\beta^*_{x,\mathbf{k}+\mathbf{q}}\beta^*_{x,\mathbf{k}'-\mathbf{q}}\beta_{x,\mathbf{k}}\beta_{x,\mathbf{k}'}\\
&+\beta^*_{y,\mathbf{k}+\mathbf{q}}\beta^*_{y,\mathbf{k}'-\mathbf{q}}\beta_{y,\mathbf{k}}\beta_{y,\mathbf{k}'}),
\end{split}
\end{equation}
where $u=U/M$, $M$ is the number of lattice sites. The nonlocal contribution is
\begin{equation}\label{eq:Vnl_CuO2}
	\begin{split}
		&V^{n}_{\mathbf{q},\mathbf{k},\mathbf{k}'}=u'(\alpha_{\mathbf{k}}\alpha_{\mathbf{k}+\mathbf{q}}[\beta_{x,\mathbf{k}'}\beta^*_{x,\mathbf{k}'-\mathbf{q}}(1+e^{-iq_x})\\
		&+\beta_{y,\mathbf{k}'}\beta^*_{y,\mathbf{k}'-\mathbf{q}}(1+e^{-iq_y})]
+\alpha_{\mathbf{k}'}\alpha_{\mathbf{k}'-\mathbf{q}}\\
&[\beta_{x,\mathbf{k}}\beta^*_{x,\mathbf{k}+\mathbf{q}}(1+e^{iq_x})+\beta_{y,\mathbf{k}}\beta^*_{y,\mathbf{k}+\mathbf{q}}(1+e^{iq_y})]),
	\end{split}
\end{equation}
where $u'=U'/M$. For $\mathbf{q}=0$, $U'=0$, and $t_0/\epsilon_d\ll1$, we obtain a simple analytical expression for the dependence on the wavevector,
$$
\frac{V_{0,\mathbf{k},\mathbf{k}'}}{u}=1+\frac{4t_0^4}{\epsilon_d^4}
[\sin^2\frac{k_x}{2}\sin^2\frac{k'_x}{2}+\sin^2\frac{k_y}{2}\sin^2\frac{k'_y}{2}].
$$
This matrix element is minimized for wavevectors on neighboring BZ edges, $|k_x|=\pi$, $|k'_y|=\pi$ or vice versa. This degeneracy is lifted by the nonlocal contribution, resulting in a minimum at $\mathbf{k}=\mathbf{k}_1$, $\mathbf{k}'=\mathbf{k}_2$ [Fig.~\ref{fig:CuO}(b)], consistent with the symmetry analysis in Section~\ref{sec:lattice}. For $u'=0.1u$, $t_0=0.5\epsilon_d$ estimated from the dispersion in cuprates~\cite{Kowalski2021,Weber2012}, $V_{\mathbf{k},\mathbf{k}}/u=0.64$ for $\mathbf{k}$, $\mathbf{k}'$ at the same $X$-point, $14\%$ larger than the minimum of $0.56$ for $\mathbf{k}$ and $\mathbf{k}'$ at different $X$-points. The difference increases with decreasing $\epsilon_d$, which may be related to its inverse relation with $T_c$~\cite{Weber2012}. Note that these effects are absent in the commonly used reduced Hubbard models of cuprates projected on Cu.

\begin{figure}
	\centering
	\includegraphics[width=1.0\columnwidth]{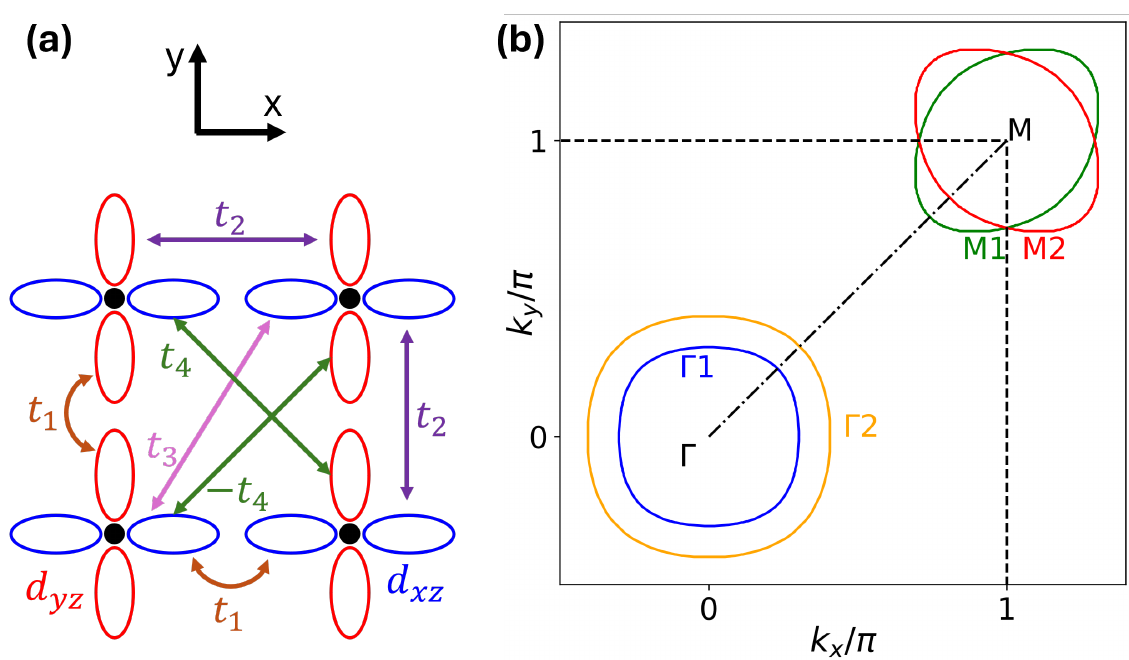}
	\vspace{-5pt}
	\caption{\label{fig:multiband_Fermi} (a) Schematic of hopping parameters for the 2-orbital model on a square lattice. (b) Fermi surface in the folded BZ replicating electron and hole pockets of undoped Fe-based superconductors.}
\end{figure}

\section{Interaction on the bipartite lattice}\label{sec:bipartite}

In this section, we consider a bipartite lattice approximating the 2D plane of Fe-based superconductors~\cite{raghu2008minimal,moreo2009properties,graser2009near,ikeda2010phase,fernandes2016low}. A minimal tight-binding model that reproduces the essential features of their band structure includes two orbitals ($d_{xz}$ and $d_{yz}$) and hopping up to the second-nearest neighbor [Fig.~\ref{fig:multiband_Fermi}(a)]~\cite{raghu2008minimal,moreo2009properties},
\begin{equation}\label{eq:Hk_Fe}	
\begin{split}
	\hat{H}_k=-\sum_{\mathbf{n},\mathbf{l_x},\mathbf{l_y},s}[
    t_1(\hat{d}_{\mathbf{n},xz,s}^+\hat{d}_{\mathbf{n}+\mathbf{l_x},xz,s}+\hat{d}_{\mathbf{n},yz,s}^+\hat{d}_{\mathbf{n}+\mathbf{l_y},yz,s})\\
    +t_2(\hat{d}_{\mathbf{n},xz,s}^+\hat{d}_{\mathbf{n}+\mathbf{l_y},xz,s}+\hat{d}_{\mathbf{n},yz,s}^+\hat{d}_{\mathbf{n}+\mathbf{l_x},yz,s})\\
    +t_3(\hat{d}_{\mathbf{n},xz,s}^+\hat{d}_{\mathbf{n}+\mathbf{l_x}+\mathbf{l_y},xz,s}+\hat{d}_{\mathbf{n},yz,s}^+\hat{d}_{\mathbf{n}+\mathbf{l_x}+\mathbf{l_y},yz,s})\\
	+i^{l_x+l_y}t_4(\hat{d}_{\mathbf{n},xz,s}^+\hat{d}_{\mathbf{n}+\mathbf{l_x}+\mathbf{l_y},yz,s}+\hat{d}_{\mathbf{n},yz,s}^+\hat{d}_{\mathbf{n}+\mathbf{l_x}+\mathbf{l_y},xz,s})+h.c]\\
    +\sum_{\mathbf{n},s}\epsilon_d(\hat{d}^+_{\mathbf{n},xz,s}\hat{d}_{\mathbf{n},xz,s}+\hat{d}^+_{\mathbf{n},yz,s}\hat{d}_{\mathbf{n},yz,s}),
\end{split}
\end{equation}
where $\hat{d}_{xz/yz}$ is the particle operator for $d_{xz}$ and $d_{yz}$ orbitals and $\mathbf{l_x}$ ($\mathbf{l_y}$) is a unit vector in the x (y) direction. The bipartite nature is accounted for by the BZ folding to the 2-Fe cell. Using the tight-binding parameter values $t_2=-1.3~t_1$, $t_3=t_4=0.85~t_1$, and $\epsilon_d=-1.45~t_1$ estimated for Fe-based superconductors~\cite{raghu2008minimal,moreo2009properties}, this model reproduces the two electron pockets ($M1$ and $M2$) at the $M$ point, and two hole pockets ($\Gamma1$ and $\Gamma2$) at the $\Gamma$-point of the folded BZ, Fig.\ref{fig:multiband_Fermi}(b)~\cite{raghu2008minimal}. 

\begin{figure}
	\centering
	\includegraphics[width=1.0\columnwidth]{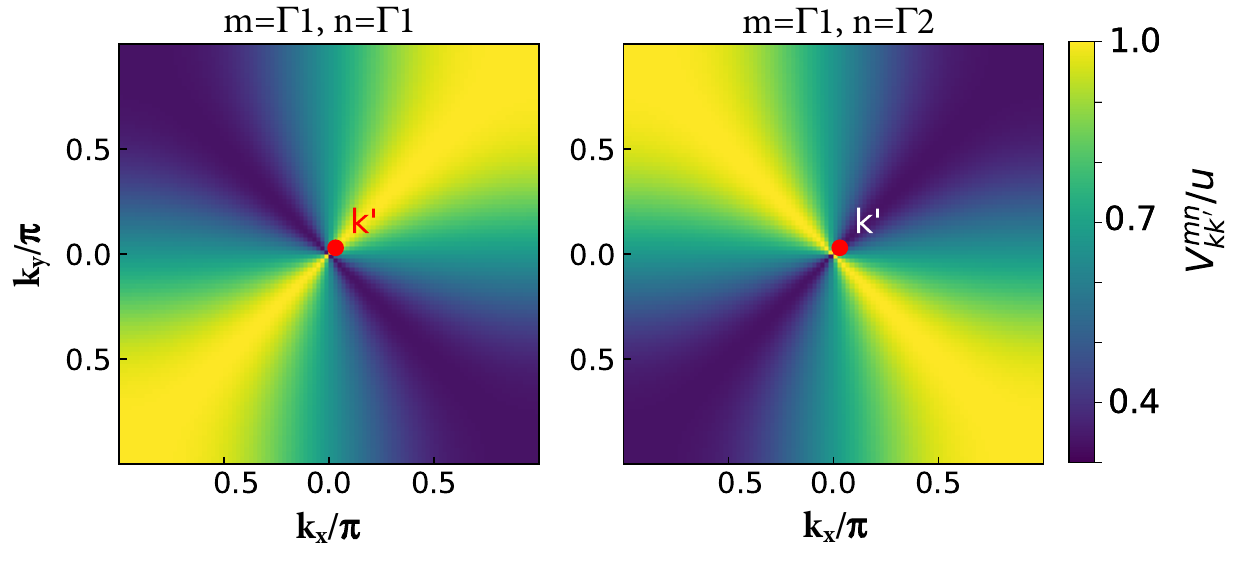}
	\vspace{-5pt}
	\caption{\label{fig:Vkk_Gamma_multiband} $V_{\mathbf{k},\mathbf{k}',\mathbf{q}}/u$ vs $\mathbf{k}$ for a bipartite square lattice, with $\mathbf{k}'=(\delta,\delta)$, $\mathbf{q}=0$, and model parameters as discussed in the text. Left: states in the same band, right: in different bands.}
\end{figure}

We use the Mott-Hund's multi-orbital interaction Hamiltonian~\cite{graser2009near,ikeda2010phase,fernandes2016low}
\begin{equation}\label{eq:int_Fe}
    \begin{split}
		\hat{H}_{int}&=U\sum_{\mathbf{n},\mu}\hat{n}_{\mathbf{n},\mu,\uparrow}\hat{n}_{\mathbf{n},\mu,\downarrow}\\
		&+U'\sum_{\mathbf{n},\mu < \nu}(\hat{n}_{\mathbf{n},\mu,\uparrow}+\hat{n}_{\mathbf{n},\mu,\downarrow})(\hat{n}_{\mathbf{n},\nu,\uparrow}+\hat{n}_{\mathbf{n},\nu,\downarrow})
        \\
		&+J\sum_{\mathbf{n},\mu < \nu} \sum_{\sigma,\sigma'}\hat{d}_{\mathbf{n},\mu,\sigma}^+\hat{d}_{\mathbf{n},\nu,\sigma'}^+\hat{d}_{\mathbf{n},\mu,\sigma'}\hat{d}_{\mathbf{n},\nu,\sigma}\\
        &+J'\sum_{\mathbf{n},\mu < \nu} \hat{d}_{\mathbf{n},\mu,\uparrow}^+\hat{d}_{\mathbf{n},\nu,\downarrow}^+\hat{d}_{\mathbf{n},\mu,\uparrow}\hat{d}_{\mathbf{n},\nu,\downarrow},
	\end{split}
\end{equation}
where $\mu$, $\nu$ are the orbital indices. The parameters are related by $U' = U-2J$ and $J'=J$, and $J=U/6$ for Fe-pnictides~\cite{ikeda2010phase,fernandes2016low}. The interaction matrix element is  
\begin{equation}\label{eq:Vint_multiband}
    \begin{split}
        V_{\mathbf{q,k,k'}}^{m,n} &= \langle0|\hat{c}_{\mathbf{k}'-    \mathbf{q},n,\downarrow}\hat{c}_{\mathbf{k}+\mathbf{q},m,\uparrow}\hat{H}_{int}\hat{c}^+_{\mathbf{k},m,\uparrow}\hat{c}^+_{\mathbf{k}',n,\downarrow}|0\rangle
        \\
        &=u\sum_\mu \alpha_{n,\mu,\mathbf{k'-q}}^*\alpha_{m,\mu,\mathbf{k+q}}^*\alpha_{m,\mu,\mathbf{k}}\alpha_{n,\mu,\mathbf{k'}}\\
        &+u\sum_{\mu<\nu}[
        \frac{2}{3}\alpha_{n,\nu,\mathbf{k'-q}}^*\alpha_{m,\mu,\mathbf{k+q}}^*\alpha_{m,\mu,\mathbf{k}}\alpha_{n,\nu,\mathbf{k'}}\\
        &+ \frac{1}{6}\alpha_{n,\nu,\mathbf{k'-q}}^*\alpha_{m,\mu,\mathbf{k+q}}^*\alpha_{m,\nu,\mathbf{k}}\alpha_{n,\mu,\mathbf{k'}} \\
        &+ \frac{1}{3}\alpha_{n,\nu,\mathbf{k'-q}}^*\alpha_{m,\nu,\mathbf{k+q}}^*\alpha_{m,\mu,\mathbf{k}}\alpha_{n,\mu,\mathbf{k'}}
        ],
    \end{split}
\end{equation}
where $\alpha_{m,\mu,\mathbf{k}}$ is the amplitude of the $m$-th band on orbital $\mu$. The electron and the hole bands are degenerate at the high-symmetry points. To avoid ambiguity stemming from this
degeneracy, we use a small displacement from $\Gamma$ and $M$ points, $\mathbf{k}'=(\delta,\delta)$ for the hole pockets, and $\mathbf{k}'=(\pi-\delta,\pi-\delta)$ for the electron pockets, where $\delta<<1$. 

\begin{figure}
	\centering
	\includegraphics[width=1.0\columnwidth]{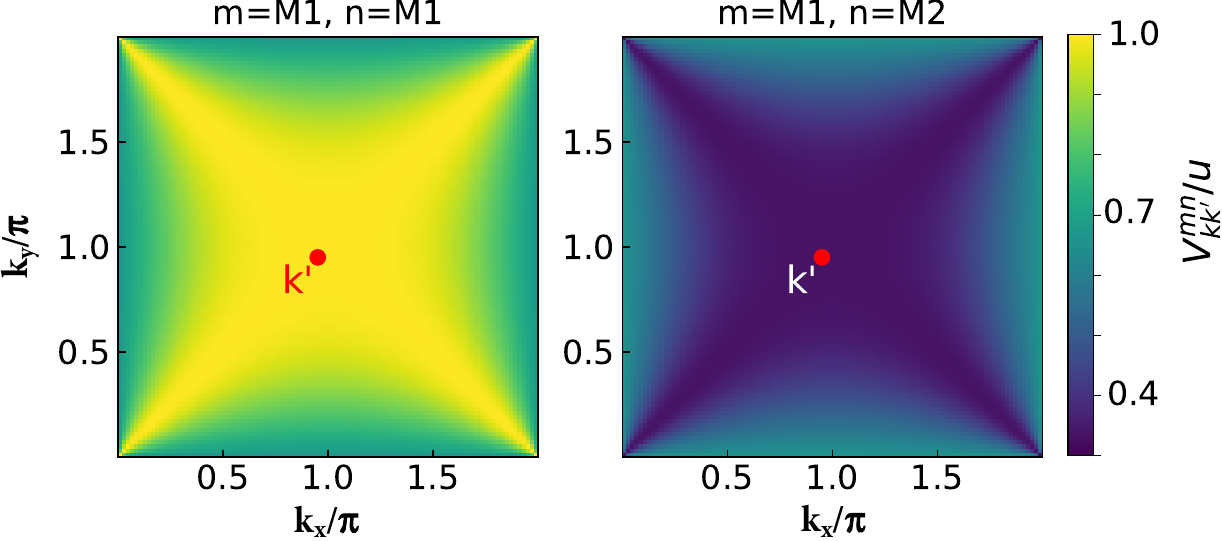}
	\vspace{-5pt}
	\caption{\label{fig:Vkk_M_multiband} $V_{\mathbf{k},\mathbf{k}',\mathbf{q}}/u$ vs $\mathbf{k}$ for a bipartite square lattice, with $\mathbf{k}'=(\pi-\delta,\pi-\delta)$, $\mathbf{q}=0$, and model parameters as discussed in the text. Left: states in the same band, right: in different bands.}
\end{figure}

Studies of the effects of doping, and high $T_c$ observed in FeSe where the hole pockets are absent~\cite{PhysRevLett.117.117001,Kreisel2020} suggest that the presence of both hole and electron pockets is not essential for pairing. Interactions within electron or hole pockets involve small momentum transfer $\mathbf{q}\approx0$. For the hole pocket, the $\Gamma$- point is a singular saddle point of the $\mathbf{q}=0$ matrix element, while the M-points are minima or maxima, as determined by the direction of $\mathbf{k}'$ [Fig.~\ref{fig:Vkk_Gamma_multiband}].  At small $\mathbf{k}$ the interaction energy varies with wavevector direction from the minimum of $0.33 u$ along one diagonal of BZ, to the maximum of $1 u$ along another diagonal, with the opposite dependences for inter- and intra-band interactions. These anisotropic dependences are associated with the anisotropy of the orbital composition of band states, which alternates between predominantly $d_{xz}$ and $d_{yz}$ character upon $90^\circ$ rotation. For $\mathbf{k}'$ near the M-point, the interaction energy reaches a maximum 
of $1u$ at the $\Gamma$- and M-points for the same band, and a minimum of $0.33u$ for different bands [Fig.~\ref{fig:Vkk_M_multiband}]. All these behaviors are consistent with the symmetries of wavevector dependence discussed in Section~\ref{sec:lattice}, including the fact that both the minimum and the maximum of interaction energy are reached, for different band combinations, at the high-symmetry $\Gamma$- and $M$-points [see top right inset in Fig.~\ref{fig:Umklapp}(b)].

\section{Effects of wavevector-dependent interaction on pairing} \label{sec:pairing}

We analyze the possible role of wavevector-dependent interactions on pairing, which is assumed to be facilitated by an additional attraction mechanism likely of electronic origin such as Kohn-Luttinger overscreening~\cite{PIMENOV2022169049}. Our assumption is that this attractive contribution is generally weaker than the Coulomb repulsion, so the latter must be minimized to achieve a net attraction. This is different from BCS, where retarded attraction results from the different timescales of electron and phonon dynamics~\cite{tinkham2004introduction}.

In the Hubbard models of SC in cuprates, pairing is commonly described by the real-space nearest-neighbor resonating valence bond (RVB)-like d-wave singlet correlations~\cite{Kowalski2021,Xu2024}
\begin{equation}\label{eq:Delta}
	\begin{split}
		\Delta=\frac{1}{M}\langle \sum_\mathbf{n}\hat{c}_{\mathbf{n},\uparrow}\hat{c}_{\mathbf{n}+\hat{x},\downarrow}+\hat{c}_{\mathbf{n}+\hat{x},\uparrow}\hat{c}_{\mathbf{n},\downarrow}\\
		-\hat{c}_{\mathbf{n},\uparrow}\hat{c}_{\mathbf{n}+\hat{y},\downarrow}-\hat{c}_{\mathbf{n}+\hat{y},\uparrow}\hat{c}_{\mathbf{n},\downarrow}\rangle
	\end{split}
\end{equation}
even though the existence of a well-defined Fermi surface in cuprates is inconsistent with the RVB model itself~\cite{PhysRevB.35.8865,PhysRevB.46.5621}. Here, $\hat{c}_{\mathbf{n},s}$ is the onsite particle operator in the projected basis. In reciprocal space, 
\begin{equation}\label{eq:RVB_recipr}
\Delta=\sum_k\Delta_k=\frac{1}{M}\langle \sum_\mathbf{k}(\cos k_x-\cos k_y)\hat{c}_{\mathbf{k},\uparrow}\hat{c}_{-\mathbf{k},\downarrow}\rangle.
\end{equation}

The interactions are scaled by the density of states in the interaction matrix
\begin{equation}\label{eq:int_mat}
	g(\mathbf{k},\mathbf{k})=V_{\mathbf{k},-\mathbf{k},\mathbf{k}'-\mathbf{k}}/\sqrt{v_F(\mathbf{k})v_F(\mathbf{k}')},
\end{equation}
where $v_F(\mathbf{k})$ is the Fermi velocity minimized close to the van Hove singularities at the $X$-points~\cite{Markiewicz2023}. The interaction energy can then be approximated by two contributions, $V_{\mathbf{k},-\mathbf{k},0}$, $V_{\mathbf{k},-\mathbf{k},\mathbf{q}_0}$, with $\mathbf{k}\approx\mathbf{k}_1$ or $\mathbf{k}_2$ describing interactions of pairs near the same and different $X$-points, respectively. Based on Eq.~(\ref{eq:RVB_recipr}), it can be evaluated as the energy of the state
\begin{equation}\label{eq:Cooper}
	\psi_C(\mathbf{k})=\frac{1}{\sqrt{2}}(\hat{c}_{\mathbf{k},\uparrow}\hat{c}_{-\mathbf{k},\downarrow}-
	\hat{c}_{\mathbf{k}+\mathbf{q}_0,\uparrow}\hat{c}_{-\mathbf{k}-\mathbf{q}_0,\downarrow})|0\rangle.
\end{equation}

\begin{figure}
	\centering
	\includegraphics[width=1.0\columnwidth]{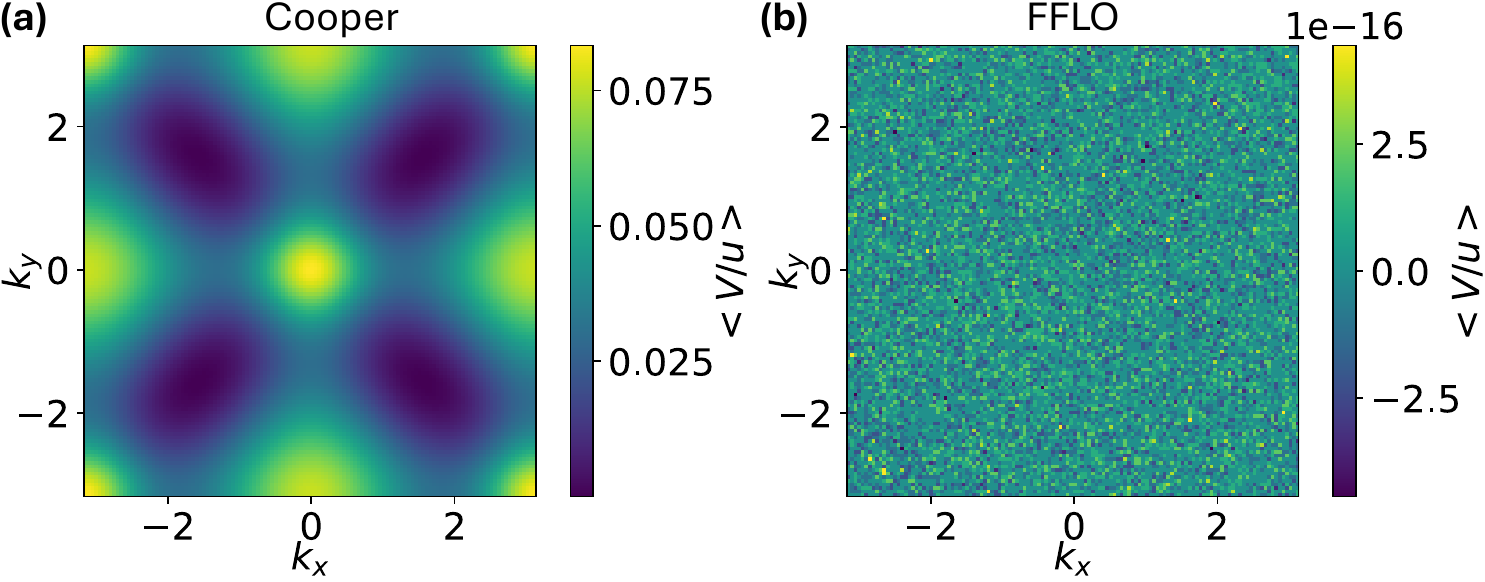}
	\vspace{-5pt}
	\caption{\label{fig:Vkk} Interaction energy normalized by $u$ for (a) superpositions of Cooper pairs with d-wave symmetry, (b) commensurate FFLO pairs with $\mathbf{Q}=(\pi/2,\pi/2)$, calculated for $u'=0.1u$, $t_0=0.5\epsilon_d$.}
\end{figure}

Figure~\ref{fig:Vkk}(a) shows the interaction energy $E_C(\mathbf{k})$ as a function of $\mathbf{k}$, for the same parameters as in Fig.~\ref{fig:CuO}(b). The characteristic repulsion energy is reduced by an order of magnitude compared to the matrix elements between Bloch states, due to the d-wave symmetry. Nevertheless, it  is maximized at the $X$-points due to the larger orbital overlap of the contribution from the same $X$-point.

Other wavevector combinations can reduce this interaction energy. Pairing between wave vectors $\mathbf{k}$ and $\mathbf{q}_0-\mathbf{k}$ describes commensurate FFLO state characterized by the wave vector $\mathbf{Q}=(\pi/2,\pi/2)$~\cite{Casalbuoni2004}. A superposition of such pair wave functions with d-wave symmetry,
\begin{equation}\label{eq:FFLO}
	\psi_{FFLO}(\mathbf{k})=\frac{1}{\sqrt{2}}(\hat{c}_{\mathbf{k},\uparrow}\hat{c}_{-\mathbf{k}+\mathbf{q}_0,\downarrow}-
	\hat{c}_{\mathbf{k}+\mathbf{q}_0,\uparrow}\hat{c}_{-\mathbf{k},\downarrow})|0\rangle
\end{equation}
has a vanishing interaction energy, Fig.~\ref{fig:Vkk}(b). 


A bosonic field carrying momentum $(\pi,\pi)$ was assumed in several models of cuprates~\cite{PhysRevB.55.3173,PhysRevB.79.134512, PhysRevLett.88.257001}, but there is presently no direct evidence for the FFLO state. We now show that this state is related to both residual AF ordering believed to exist in superconducting cuprates~\cite{Armitage2010,Boschini2020} and zero-momentum Cooper pairing. We introduce operators
\begin{equation}\label{eq:fermions}
	\begin{split}
		\hat{a}_{\mathbf{k},s}=\frac{1}{\sqrt{2}}(\hat{c}_{\mathbf{k},s}+\hat{c}_{\mathbf{k}+\mathbf{q}_0,s}),\\
		\hat{b}_{\mathbf{k},s}=\frac{1}{\sqrt{2}}(\hat{c}_{\mathbf{k},s}-\hat{c}_{\mathbf{k}+\mathbf{q}_0,s})
	\end{split}
\end{equation}
defining two fermions that can be interpreted as opposite pseudo-spins on the folded BZ $|k_x|+|k_y|\le\pi$. 
Using the same parameters as in Section~\ref{sec:cuprates}, the repulsion energy between opposite pseudo-spins vanishes if nonlocal interactions are neglected, and is about 40 times smaller than repulsion between the same pseudo-spins in the extended Hubbard model. The AF-ordered state is
\begin{equation}\label{eq:AF}
	\psi_{AF}=\prod_{\mathbf{k}}\hat{a}^+_{\mathbf{k},s}\hat{b}^+_{\mathbf{k},-s}|0\rangle,
\end{equation}
where the direction of $s$ defines the Neel vector. If one neglects repulsion between opposite pseudo-spins, the interaction energy vanishes in this state. This energy gain is offset by the kinetic energy cost of mixing between states with different single-particle energies. This competition can limit ordering described by Eq.~(\ref{eq:AF}) to the reciprocal space regions near the $X$-points where the band energies are close, which describes residual AF ordering and may explain the spectral broadening of single-particle dispersion near these points~\cite{Armitage2010}.

The pair wavefunction describing such residual AF ordering in the two-particle limit,  $\psi=\hat{a}^+_{\mathbf{k},\uparrow}\hat{b}^+_{-\mathbf{k},\downarrow}|0\rangle$, can be written as a superposition of a d-wave Cooper pair $\psi_C$ and the FFLO pair, 
\begin{equation}\label{eq:psi01}
	\begin{split}
		\psi=\frac{1}{2}[(\hat{c}^+_{\mathbf{k},\uparrow}\hat{c}^+_{-\mathbf{k},\downarrow}-\hat{c}^+_{\mathbf{k}+\mathbf{q}_0,\uparrow}\hat{c}^+_{-\mathbf{k}-\mathbf{q}_0,\downarrow})\\
		-(\hat{c}^+_{\mathbf{k},\uparrow}\hat{c}^+_{-\mathbf{k}-\mathbf{q}_0,\downarrow}-\hat{c}^+_{\mathbf{k}+\mathbf{q}_0,\uparrow}\hat{c}^+_{-\mathbf{k},\downarrow})]|0\rangle,
	\end{split}
\end{equation}
The possibility of such pairing is supported by the observation of ``hot spots" $-$ regions with enhanced superconducting gap and decreased electronic spectral weight around the points of intersection between Fermi surface and AF BZ boundary~\cite{Armitage2010}. In these regions, both the kinetic and the interaction energies are minimized for the FFLO pairing. The latter mixes different single-electron momenta resulting in reduced spectral weight. 

For the bipartite lattice, analysis of pairing wavefunctions minimizing the interaction energy is more challenging due to the complexity of possible combinations of wavevectors and orbital states. Here, we outline the general trends inferred from the calculated matrix elements, and leave detailed analysis to future studies. At small $\mathbf{k}$ the interaction energy varies  with wavevector direction by a factor of $3$, between $0.33u$ and $1u$, Fig.~\ref{fig:Vkk_Gamma_multiband}. For $\mathbf{k}=-\mathbf{k}'$, it is  maximized for the same pocket, and minimized for different pockets. This relation reverses when the direction of $\mathbf{k}$ is orthogonal to $\mathbf{k}'$. Thus, interaction energy can be minimized by finite-momentum FFLO pairing in the same pocket, or by pairing between different pockets. Because of the different dispersions of the two pockets [see Fig.~\ref{fig:multiband_Fermi}(b)], the latter also favors FFLO state with a small pair momentum.
At the M-points, the interaction energy is minimized for pairing between different pockets, Fig.~\ref{fig:Vkk_M_multiband}. Similarly to the hole pockets, the dispersions of electron pockets are different, so the interaction energy is minimized by small-momentum FFLO pairing between different pockets. 
The sign of the pair wavefunction is reversed between the pockets at the M-points and at the $\Gamma$-point~\cite{Hirschfeld2011}, suggesting the possibility of FFLO pairing with the same wavevector $2\mathbf{Q}=(\pi,\pi)$ as discussed above for cuprates. 

\section{Summary}\label{sec:summary}

In summary, we showed that electron interactions on the lattice obey certain symmetry requirements, regardless of the underlying interaction mechanisms. The interaction energy reaches extrema for electron states formed by superpositions of wavevectors close to high-symmetry  points of the Brillouin zone. We analyzed two examples illustrating these symmetries, a square lattice with the basis AB$_2$ approximating the CuO$_2$ plane, and a bipartite square lattice approximating a 2D plane of Fe-based superconductors. In both cases, the wavevector dependence of Coulomb interaction originates from the variations of the orbital composition of the wavefunctions. In the former case, the interaction energy is minimized for combinations of wavevectors at two inequivalent $X$-points of Brillouin zone. For the bipartite lattice, a similar mechanism minimizes the interaction energy of electrons in different bands. We also investigated possible forms of pairing minimizing interaction energy, including a state with residual antiferromagnetic ordering, which combines d-wave zero-momentum Cooper pairing and a commensurate FFLO pairing. 

As a common feature of the considered models, the interaction energy is minimized for electron pairs with non-zero momentum. Finite-momentum pairing in cuprates could account for spectral broadening near the X-points and at the hot-spots where Fermi surface intersects the AF BZ boundary~\cite{Armitage2010,Norman1998,RevModPhys.75.473}, the  incoherent Cooper pairs observed above the critical temperature~\cite{zhou2019electron}, and the  pseudo-gap as a signature of such incoherent pairs. This interpretation is also supported by the properties of a class of highly resistive superconducting materials called ``bad metals"~\cite{osofsky2001new,doi:10.1021/jacs.4c06836}, where according to the Ioffe-Regel criterion the electrons are localized and therefore Cooper pair momentum is not well-defined~\cite{Hussey2004}. Similarly to HTSCs, these materials exhibit incoherent pairing above $T_c$~\cite{dubi2007nature,yang2019intermediate,zhang2022anomalous,zhang2025}.

Our analysis suggests certain symmetry criteria for unconventional SC. The common features that in the considered models play a central role are i) Fermi surface that is either localized close to the high-symmetry BZ points with maximal irreducible distance or has a large spectral weight near these points, ii) a non-trivial orbital structure of Bloch states allowing electrons to avoid each other. These features are shared by other known to us unconventional superconductors, such as multilayer graphene and transition metal dichalcogenides where the Fermi surface is formed by the pockets with distinct orbital composition localized at the corners of BZ~\cite{Chen2008,Cao2018}. The critical temperatures of quasi-2D unconventional superconductors are  limited by fluctuations due to reduced dimensionality. In 3D, the identified criteria may be satisfied by a hexagonal material characterized by heavy electron or hole pockets near the H-K line of BZ. Optimization of the 3D band structure, while more complex than in the quasi-2D unconventional superconductors, may ultimately enable room-temperature superconductivity at ambient pressure. 

We acknowledge partial support by the SEED award from the Research Corporation for Science Advancement. Bipartite lattice calculations by Y.Z. were supported by 
the subcontract No.C5808 for the US Department of Energy award No.88148 through Los Alamos National Laboratory.

\bibliographystyle{apsrev4-2}
\bibliography{HTSC}

\end{document}